# Defining a physics person: Tensions in the identity development of undergraduate women in physics

Laura M. Akesson,[1] Jessica L. Rosenberg,[1]
Nancy Holincheck,[2] and Benjamin W. Dreyfus[1]

[1]*Department of Physics and Astronomy, George Mason University, Fairfax, Virginia 22030, USA*
[2]*College of Education and Human Development, George Mason University, Fairfax, Virginia 22030, USA*

This qualitative study explores how undergraduate women define what it means to be a physics person alongside how they describe their own physics identity. Our team analyzed 120 survey responses (97%, or 116 of the 120 respondents, identified as women or non-binary) to understand what factors are associated with an individual's physics identity. Those surveyed were attendees of Conferences for Undergraduate Women in Physics, who attended 88 unique institutions in 30 states. Our participants' conceptions of what it means to be a physics person fell into two themes: (1) people who are skilled academically and in research and (2) those who have an interest in physics ranging from an enjoyable interest to a life-devoting passion for the subject. When asked to explain their own physics identities, students responded through a personal and social lens more frequently than through an academic lens. Our analysis revealed three tensions in the development of undergraduate women's physics identity: (1) the tension between identifying entirely as a physics person and having other identities that are important to who they are, (2) the tension between the need for belonging in the physics community and their view of themselves as a physics person, and (3) the tension between their expectations of a physics person and their beliefs about their own abilities or interest.

## I. INTRODUCTION

The underrepresentation of women in physics is a persistent and well-documented problem that has been addressed in the literature over the past 60 years [1–5]. Although some progress has been made over this time, physics still has one of the largest gender gaps among the sciences [6], as less than 25% of physics degrees in the United States are awarded to women. The gender gap in physics is a phenomenon that is experienced in most parts of the world [7], including, but not limited to, Africa [8], the United Arab Emirates [9], Thailand [10], Turkey [11], and Australia [12].

The gender gap in physics has been attributed to a variety of causes, including (a) masculine cultures that cultivate men's sense of belonging more than that of women, (b) insufficient early experience with computer science, engineering, and physics, and (c) gender-based differences in self-efficacy [13,14]. In the last 15 years there has been increasing interest in understanding how to build women's physics identity to support their persistence in physics majors and careers [15,16].

Identity "provides us with the tools to examine the complex mixture of the personal, social, cultural, and political aspects of what it means to become a physicist" [17]. Identity and its gendered formation have been tied to academic achievement, motivation, and intentions to stay in STEM [18–27]. It impacts an individual's ability to navigate challenges, build relationships, and find community [27–31], and it both limits and expands career considerations and, ultimately, participation and persistence in the STEM workforce [24,32–35]. Identity is an especially important consideration for women and other marginalized communities. Stets and colleagues found that the identity process is the "primary mechanism" of career choice for minority students [36].

This study is part of a larger grant-funded project examining the development of identity in undergraduate women in physics. In this paper, we detail our findings from a survey of 120 undergraduates in physics who attended 88 unique institutions across 30 U.S. states. We invited any student who identifies or had ever identified as a woman to participate.

## II. THEORETICAL FRAMEWORK

### A. Definition of identity

Our study of physics identity is framed by Gee's definition of identity as recognizing oneself and being recognized by others as a certain kind of person [37]. Gee's identity theory encompasses the kind of person one wants to be in the present as well as the kind of person one aspires to be in the future [38]. The continuous formation and motion of an individual's identity over time are described by Wenger in terms of a "learning trajectory," stating that "we define who we are by where we have been and where we are going" [39]. In both Gee's and Wenger's conceptions, identity is shaped through continued experiences and developing aspirations. The evolution of an individual's identity over time is of particular importance to our project. In this study, we recognize that undergraduate women's perceptions of themselves are formed by past experiences, current environments, and their conceptualizations of possible future selves.

### B. STEM and physics identity

Building upon Gee and Wenger's conceptions of identity [37,38], researchers define STEM and physics identities as recognizing oneself and being recognized by meaningful others as a STEM person or a physics person [15,16]. Carlone and Johnson's foundational study identified three dimensions of science identity: (a) recognition by self and others as being a STEM person, (b) one's sense of competence in STEM, and (c) one's sense of being able to perform STEM tasks (performance) [15].

A strong science identity has been linked to a higher likelihood of choosing and persisting in STEM majors, graduate programs, and careers [14,36,40]; having greater positive beliefs about their competence [18]; and viewing costs (such as hours studying) as worthwhile. The choice to leave STEM majors is unrelated to competence or ability [19], but often it is due to the culture of undergraduate science departments that fail to develop the science identity of some students [20]. Gender differences in science identity development of undergraduate students have been attributed to perceived recognition [17,21,22,41,42] and sense of belonging [19,23,43–45], and these differences begin as early as middle or high school [46]. Potvin and colleagues [47] have shown the importance of direct intervention through counternarratives in increasing women's interest in pursuing physics. Another way to increase confidence and protect students from negative experiences is through early and consistent positive external recognition [48]. For women, high STEM identity has been found to provide a buffer against negative experiences, namely sexism [24].

Carlone and Johnson's study on the science identity of successful women of color laid the groundwork for Hazari and colleagues' research exploring the role of physics identity for women in physics [15,16]. Their study led to modifications of Carlone and Johnson's model, adding the dimension of interest and combining performance and competence into a single dimension [16]. Quantitative studies of STEM identity generally treat competence and performance as a single dimension [25], but some qualitative studies have retained the distinction [33,49]. In a later study examining the experiences of first-year and senior undergraduate women in physics, Hazari and colleagues found a sense of belonging to be statistically significant for seniors, but not for freshmen [41]. This finding led to the addition of a sense of belonging as a dimension of physics identity. It supported previous work that identified a sense of belonging as an important factor in women's persistence in physics [20,43,50].

### C. Research questions

We used a qualitative survey to investigate the following research questions:

1. What are undergraduate women's conceptions of what it means to be a physics person?
2. How do undergraduate women explain the self-assessment of their own physics identities?
3. How do undergraduate women's conceptions of what it means to be a physics person relate to the self-assessment of their physics identities?

## III. METHODS

### A. Recruitment and participants

We recruited participants through the exit survey distributed to all students who attended one of the Conferences for Undergraduate Women in Physics (CUWiP) in January 2023. In 2024, the conference was renamed the Conference for Undergraduate Women and Gender Minorities in Physics (CU*iP). These regional conferences are organized annually by the American Physical Society (APS) and held concurrently in mid-January at universities around the U.S.. At CUWiP, students attend research talks by faculty, panel discussions about graduate school and careers in physics, presentations and discussions about women in physics, laboratory tours, and student research talks; they can choose to present in a student poster session. Students of all genders are welcome at CUWiP. To ensure the inclusion of non-binary and transgender students, we directed students, "If you are or ever have been a woman in STEM, we invite you to participate in our study."

Students who completed the exit survey for a January 2023 CUWiP were asked to indicate if they were willing to participate in our survey and interview study. We emailed the 367 students who indicated they were willing to participate, and a total of 120 of these students responded to the survey. Table I includes students' gender identities and race or ethnicities. The vast majority of students who responded identified as women (108 of 120). Eight identified as non-binary students, one as male, and three students chose not to share their gender identity. The survey participants included 74 White students, 26 Asian students, 10 Hispanic or Hispanic/White students, two

TABLE I. Survey participant information.

| Demographic | Category | Number of respondents |
|---|---|---|
| Race or ethnicity | White | 74 |
| | Asian, South Asian, Asian/White | 26 |
| | Hispanic, Hispanic/White | 10 |
| | Black | 2 |
| | Multiracial | 3 |
| | No response/other | 3 |
| Gender identity | Female | 108 |
| | Male | 1 |
| | Non-binary | 8 |
| | Not shared | 3 |
| Year in school | Freshman | 20 |
| | Sophomore | 24 |
| | Junior | 37 |
| | Senior | 34 |
| | Other (gap year, recently graduated) | 5 |

Black students, five multiracial students, and three who chose not to respond or identify their race. Multiracial refers to students of mixed ethnicities other than Asian/White and Hispanic/White.

The 120 survey respondents represented 30 states and 88 unique institutions. In addition to geographical diversity (see Fig. 1), the institutional characteristics also varied greatly to include public, private, large, small, an HBCU (1), HSIs (15), and community colleges (3). The scale and diversity represented in this study are a unique source of strength.

### B. Data collection

We constructed an online survey consisting of open-ended and multiple-choice questions to help us understand aspects of the participants' physics identity. This included a modified version of McDonald and colleagues' Science, Technology, Engineering, and Math (STEM) identity measure [51], replacing "STEM Person" with "Physics Person" (see Fig. 2) and using it as a proxy for physics identity. The single-item measure is a series of increasingly overlapping pairs of circles, with one circle labeled "me" and the other a "physics person". The choices ranged from A, which has no overlap between the circle for "me" and the one for "a physics person," to G, where the circles nearly completely overlap.

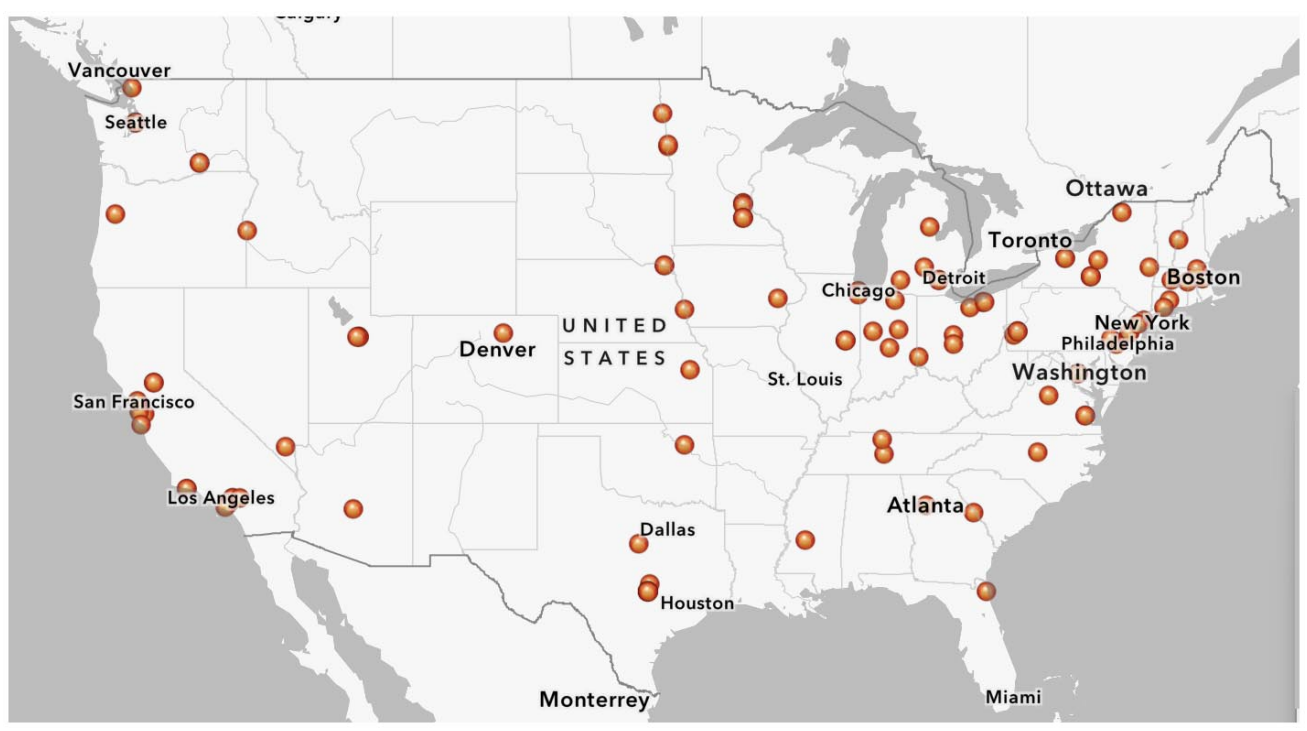

FIG. 1. The geographical location of colleges and universities attended by survey respondents.

Following the physics identity measure question, the survey contained two open-ended questions:

- Briefly explain your response to the previous question.
- What does it mean to be a physics person?

The survey went on to ask about future plans, leadership roles, mentorship, college and university information, and race/ethnicity. We placed these questions after the ones about identity to mitigate reminders of experiences or responsibilities that might have affected their answers to the identity questions.

### C. Data analysis

Data analysis of the two open-ended questions above included an initial cycle of descriptive and *in vivo* coding [52], followed by a more purposeful cycle of provisional coding. These initial rounds were completed independently by all four group members. Additional rounds of coding were done collaboratively through comparison and discussion among all four authors. This part of the process was iterative, where we refined our list of codes, clarified code definitions, and reached consensus on any discrepancies. Biases and assumptions were explicitly addressed in the group discussions.

Our initial codebook focused on STEM identity and included *a priori* codes of performance/competence, recognition, and interest. However, in our first cycle of *in vivo* coding, the research team noted that many participants described the area of their identities that did not overlap with a physics person (see Fig. 2). Students commented on their cultural and social environments and personal identities outside of physics. Other comments described their perceived progress in physics. This prompted us to add codes such as *multiple identities* and *still learning*. Similarly, we noted certain patterns in participants' responses to what it means to be a physics person, which led us to add the codes *researcher*, *wonder*, and *dedicates life*, among others. Codes that were identified four or fewer times are not included in this paper.

Our initial rounds of analysis focused on trends across the participants in our study. In our final phase of analysis, we examined each participant's data individually, looking at the relationship between their responses. Specifically, we compared how participants described their own physics identities with how they described a physics person. In our collaborative analysis, we grouped participants by the nature or level of agreement between their responses, and then we looked across our data to identify trends.

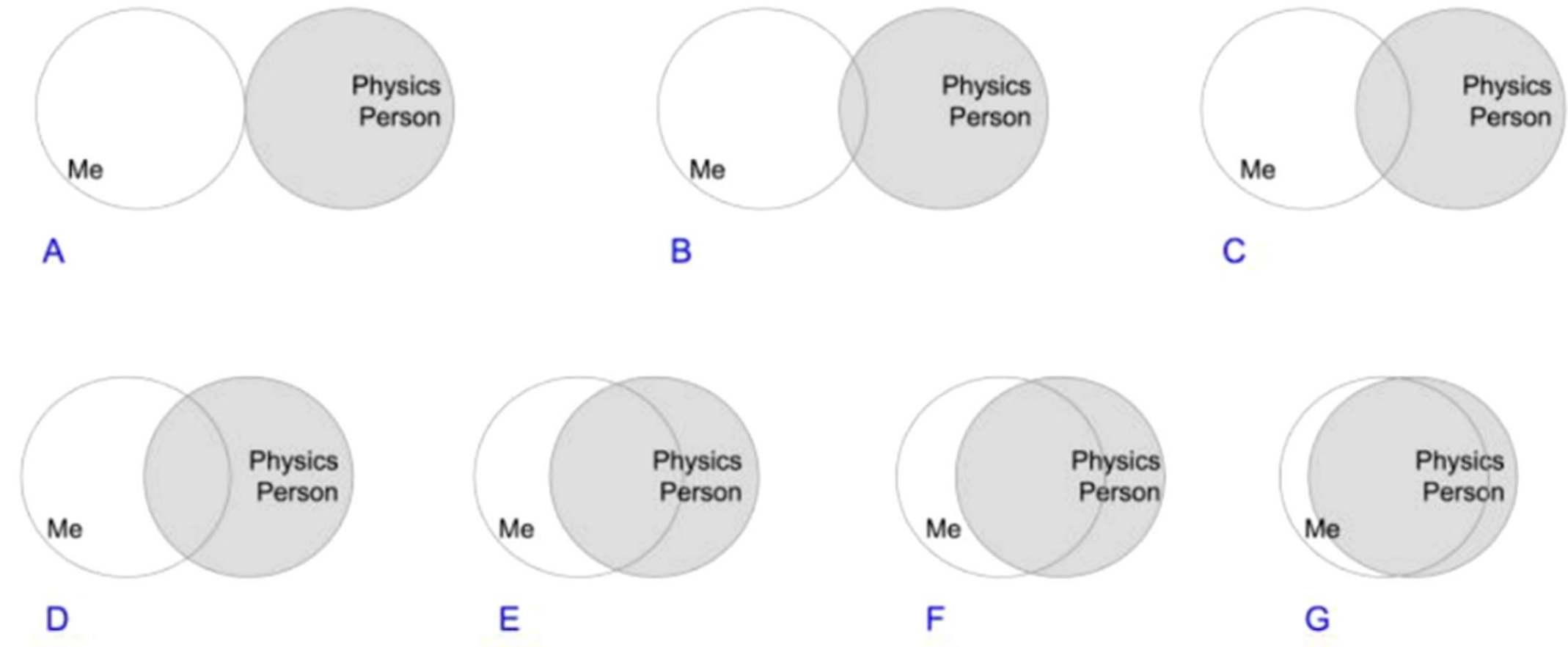

FIG. 2. A survey question and single-item measure of physics identity, based on the STEM identity measure of McDonald *et al.* [51].

## IV. FINDINGS

Due to the large number and geographic diversity of our participants' colleges and universities—including community colleges, HSIs, and BSIs—and strong presence of all undergraduate years, we are able to present a representative sample of how undergraduate physics students (1) describe a physics person and (2) explain their own physics identities. The emergent themes found within the responses include evidence of Hazari's physics identity dimensions, enhanced with greater contextual understanding due to how the students described interest, performance and competence, and sense of belonging. We note a strong influence of students' personal and social identities on their physics identities. In comparing students' descriptions of a physics person (external perspectives) with the explanations of their own physics identities (internal perspectives), we found both similarities and differences. The comparison revealed factors that have helped or hindered the development of their physics identities.

Throughout our analysis, we identify participants with a letter, A–G, and a number. The letter corresponds to the self-assessed overlap between their self-image and the image they have of a physics person (see Fig. 2). A response of "A" represents no overlap, and "G" represents a nearly complete overlap. The number identifies each student.

### A. How students described what it means to be a physics person

One of two open-ended questions on the survey we analyzed was "What does it mean to be a physics person?" Our analysis of student responses identified two predominant themes: (1) someone who enjoys and is passionate about physics, at times to the exclusion of other interests; and (2) someone who is good at physics and research, to whom concepts come naturally. Figure 3 shows the common codes and their frequency. Codes that were identified four or fewer times are not included in this paper.

#### *1. Interest, wonder, passion, and dedicates life*

The first theme includes responses coded with *interest*, *wonder*, *passion*, and *dedicates life*. This mixture of *in vivo* and provisional codes was grouped together as components on a sliding scale of Hazari *et al.*'s interest dimension [16]. The scale is a continuous spectrum ranging from innocuous enjoyment to constant wonder, emotional passion, and singular rumination and focus. Students who indicated that a physics person is interested in physics content, doing physics, enjoys doing or studying physics were coded with *interest*. When a student's definition included that a physics person feels strong emotions for physics, using words such as "love" or "passion," responses were coded with *passion*. Students who described physics people with a quality about them that demonstrates curiosity, excitement, and wonder about the world and/or universe, and how physics explains or fits in were coded with *wonder*. Lastly, responses that contained a sense that to be a physics person, physics needs to be one's overarching focus of life, were coded as *dedicates life*.

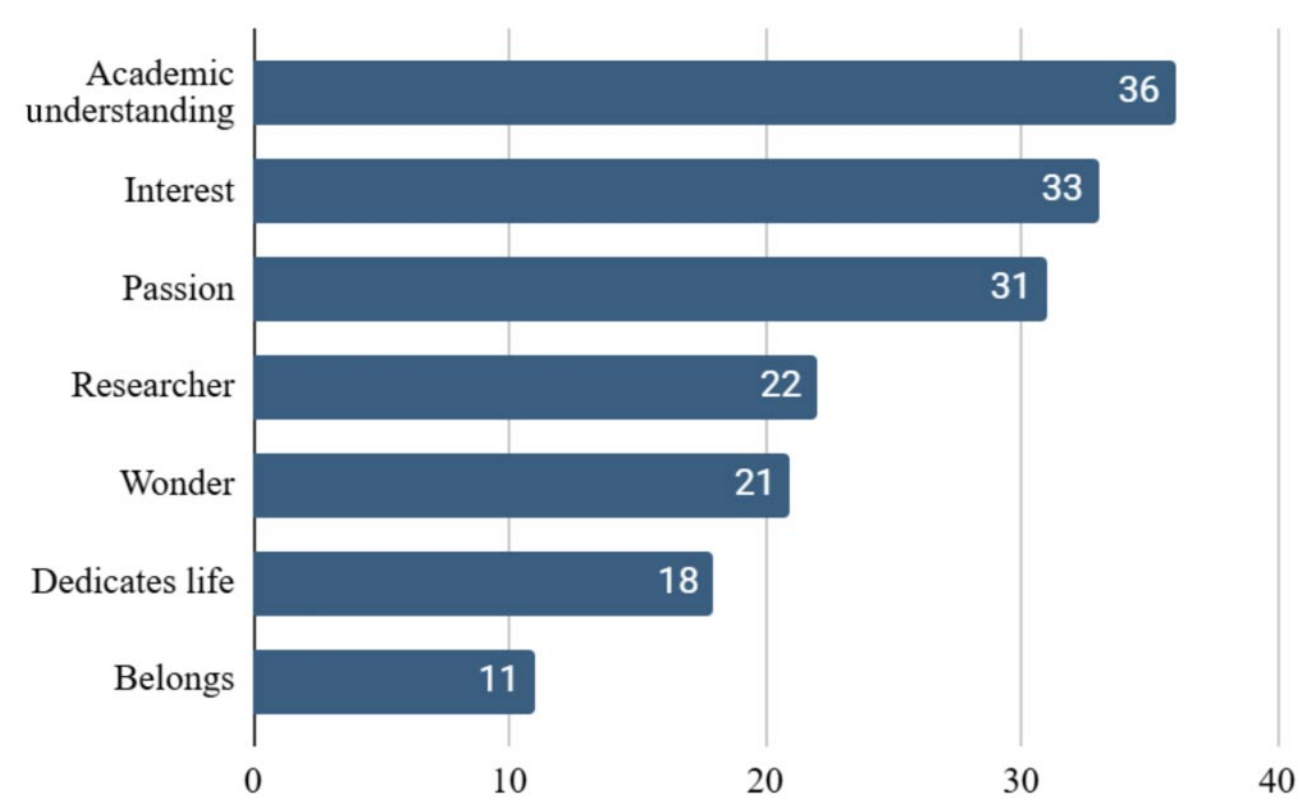

FIG. 3. Graph of the frequencies of the coded student responses to the survey question, "What does it mean to be a physics person?" When a student's open-ended responses included several characteristics of a physics person, multiple codes were applied. Codes that were identified four or fewer times are not included in this graph or paper.

Many students used the word "interest" verbatim and defined a physics person as "someone who is interested in physics" [F9]. Others used phrases that aligned with our definition of interest, describing a physics person as one "who enjoys and respects the field of physics, someone who is always willing to learn about new physics" [G5]. Interest itself was described in many ways, as being "genuine," "comfortable," and "inherent," causing people to pursue physics "for the sake of learning" [E28] and "just for fun" [E44].

*Passion* was a common *in vivo* code, with a physics person defined as "one who is deeply involved in the field…and is deeply passionate about the field" [E14]. Students described this passion as a deep emotion, enthusiasm, and love for physics, and directed toward "understanding nature's inner workings" [E21].

*Wonder* defined a physics person who "has endless curiosity about the world" [D12] and the "nitty gritty details" [B7] of "the interactions of things and systems… (physical, electrical, etc.)" [D24]. In their words, physics people are naturally drawn and committed to this deep curiosity.

At first glance, this definition of a physics person is joyful, positive, and affirming. Yet, toward the latter end of the spectrum, the responses evolved from healthy interest to deep passion to all-encompassing obsession. A number of responses contained the idea that a physics person *dedicates* one's *life* to physics, "spending a significant portion of your time on related activities, taking steps that keep you spending a significant portion of your time on related activities" [F15]. Students at this end of the sliding scale defined a physics person as one "who will put physics over everything else in their life. They prioritize it over money, over choosing where to live, etc." [E10]. Physics people are meant to center not only their professional time and energy on physics but also their social identity as well. Another student's entire response was, "I see physics people as people [whose] whole life revolves around their work in physics, and a large part of their community and identity is centered in physics" [E1].

#### *2. Academic understanding and researcher*

The second conception of a physics person is someone who has a high level of academic understanding and applies this aptitude to research. Responses that expressed this understanding of physics and academic work, commonly referencing intrinsic/natural abilities, were coded as *academic understanding*. Students defined a physics person as being "good at physics" [C6, E6, E22, E31, F2, F10] with "good math and computation skills" [B7]. Many wrote that physics people have an intuition for the subject, have a natural capacity, and can "easily do physics problems" [C3]; that "your brain understands how to do physics" [E18]; and that one is "able to *naturally* draw correlations and connections between concepts very *easily*, know how to apply math skills to physics situations" [D1]. Some included popular generalities that "they also tend to be a typical 'nerd'," [D6] and "tends to be Albert Einstein or someone like that" [D25]. Only one student expressed the idea that "understanding does not have to be perfect" [F11]. The code of *academic understanding* and the idea that a physics person possesses innate ability were prevalent across all undergraduate years.

When students defined a physics person as someone who applies their computational and problem-solving abilities in *research*, students described physics people as those who are involved, capable, and successful researchers. They are "dedicated to their studies or research and are highly intelligent" [D6], are "in the middle of scientific research and discoveries" [B1], and are "someone who can analyze a given situation/setup through the laws of physics at any given time, and are involved in physics research" [D4]. Students also expressed the idea of enjoyment in expanding understanding, that a physics person "likes lab work, research, or engineering" [F16] and "is constantly challenging the currently accepted science through thought, research, and study. They are problem solvers and are not afraid to fail repeatedly on their path to knowledge" [F12]. Within the responses, some students acknowledged a researcher as a physics trope, with one citing, "…that's the stereotypical definition of a physics person" [D19], or adding it as a characteristic that should be present: "I guess research too?" [D5].

One additional code that did not fit within either theme described above was sense of *belonging*. Only 11 of the 160 applied codes reflected the idea that being involved in, contributing to, and feeling a part of the physics *community* is an aspect that defines a physics person.

### B. Students' categorization of their physics identities

As described in the methods section, the survey question addressing participants' self-image as a physics person had two parts: a lettered response to the single-item measure (See Fig. 2), which we used as a proxy for physics identity and an open-ended explanation of their responses. We use these two responses to better understand how students see themselves as physics people.

Figure 4 shows the distribution of 120 survey respondents. A total of 18 students (15.0%) chose A, B, or C, a low physics identity. The remaining 102 students reported mid (D or E, $n = 75$, 62.5%) or high (F or G, $n = 27$, 22.5%) physics identity. We found the majority of students (85.0%) classified themselves in the range of mid-to-high physics identities. We note that these participants were attendees of CUWiP who agreed to take part in our study. Voluntarily attending this conference suggests a curiosity or an interest in a community centering undergraduate women in physics, or already having an identity aligned with this community. Attendance might also indicate support and recognition of the student by their university or college physics department. Considering our population, one might expect a shift toward higher physics identity responses relative to the overall population.

Also included in Fig. 4 are colors representing the year in school (freshman to senior, and other). Interestingly, physics identity responses (A–G) were not significantly related to the year in school. Juniors and seniors who have presumably spent more time in the physics community, continued to choose physics over multiple semesters, completed more coursework, and declared majors, were not more likely to have a greater physics identity.

### C. How students explained their physics identity

When students were asked to explain their physics identities(their lettered responses), many included responses that corresponded to Hazari's physics identity dimensions of interest and performance and competence. The majority of students chose to explain their physics identity in the context of their social and personal identities.

The eight most frequent coded responses with their number of occurrences are shown in Fig. 5. Out of the responses, 42.5% were assigned more than one code. Most codes contain both negative and positive descriptions, meaning that a response reflecting a lack of belonging and another representing a strong sense of belonging were both coded with belonging.

#### *1. Academic association with physics content: Interest, performance, and competence*

Many students (47 of 120, 39%) described their identity through an academic association with the subject matter of physics and their sense of their ability to engage with the content. These responses included expressions of *interest* or curiosity about learning, doing, and applying physics concepts, and beliefs about their ability to *understand* physics content (*competence*) and *perform* physics tasks (*performance*).

Expressions of interest in physics were present across all levels of physics identity (A–G). Interest was described with words of "love" and "enjoyment" for the content of the subject matter, "I love physics and I'm interested in it…" [C3] and "I enjoy studying physics and understand the world around me better for having studied it" [D24]. Students were interested in a "broad range" of various topics, were enthusiastic and "passionate about physics" [C4], and "really enjoy[ed] learning about physics" [E13]. They were eager to talk to others and spent a lot of time "even outside of classes" [F13], "talk[ing] about physics to my housemates, and spend[ing] much of my time thinking about physics" [F15], and "studying and reading physics" [G7].

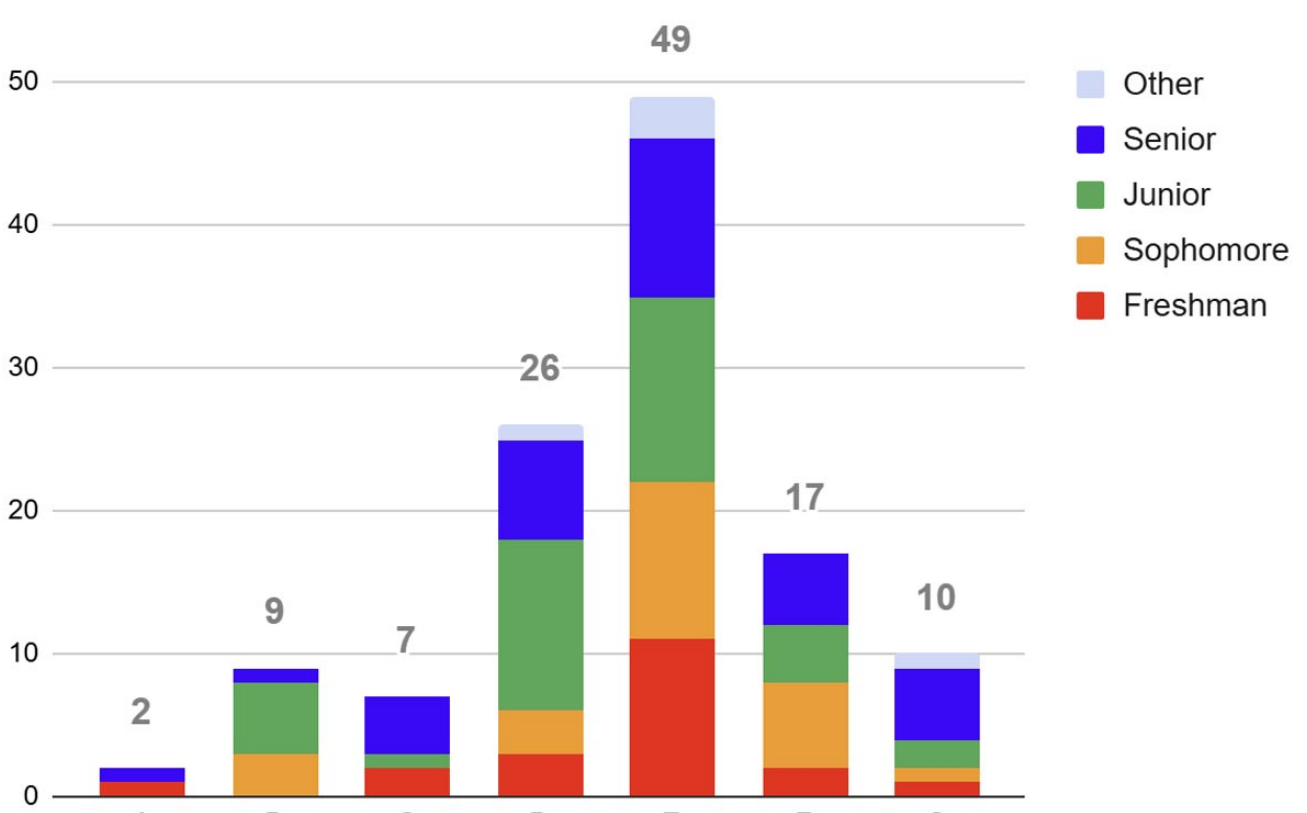

FIG. 4. Distribution of how students ($n = 120$) categorize their physics identity (A is the lowest, G is highest). See Fig. 2. Student participants from across the country were recruited from Conferences for Undergraduate Women in Physics (CUWiP) in 2023. The color bands represent the undergraduate year of the students.

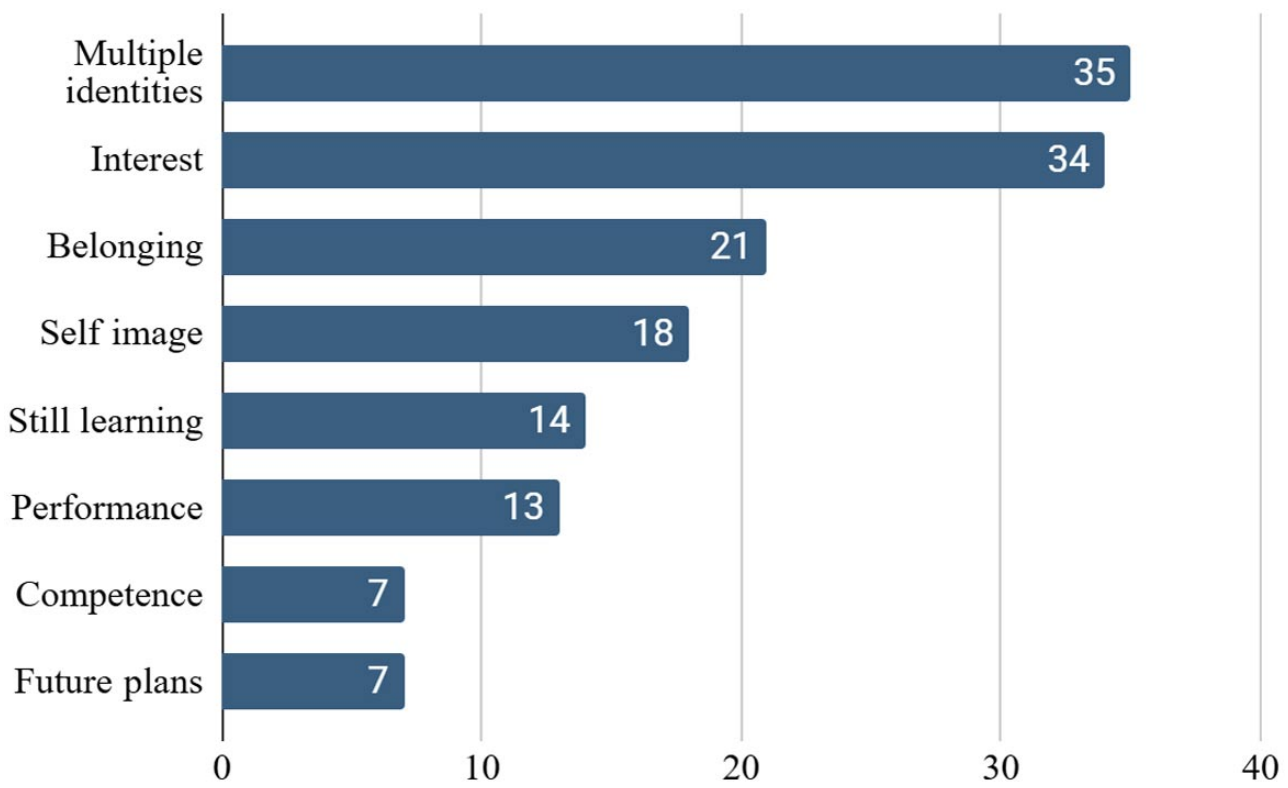

FIG. 5. Graph of the frequencies of the coded responses to the students' descriptions of their own physics identities (A—G responses, see Figs. 2 and 4). When a student's open-ended responses included several characteristics of a physics person, multiple codes were applied. Codes that were identified four or fewer times are not included in this graph or paper.

Although interest was a common component in participants' descriptions of physics identity, it was frequently paired with qualifying statements. These statements included the idea of falling short of a perceived physics standard. One student explained how she felt challenged by the speed of coursework: "I love physics…but it moved too quickly for me and so I never fully became a 'physics person'" [C3]. Another cited her lack of research experience and insufficient memory: "I enjoy studying physics…however, because I don't participate in research and because I don't recall all of my training, I don't really consider myself to be a physicist" [D24]. A few responses compared themselves to their peers and found themselves lacking: "I do love physics, and I think I have an aptitude for it, but I started studying late, and don't feel much of a kindred spirit to my peers. Maybe not nerdy enough. Not amazing at math. Not in love with school" [C7]. Even those with higher physics identities paired their interest with self-doubt, stating they were "very interested in physics but not that good in it" [E40]. After a lengthy response describing their love of physics, its "problem-solving and logical thinking," and talking about and spending time thinking about physics, a student still remained unsure about her future in physics: "It's just that the next obvious step on the physics path is not one I'm sure I want to take" [F15].

Another set of qualifying statements that were combined with interest denoted that physics was only part of a more complex identity: "I enjoy physics but I also enjoy many other subjects and activities" [D26]. This emergent topic of multiple identities was common and is addressed later in the paper.

Of the 18 participants who wrote about *performance* and/or *competence*, two students made positive comments regarding their performance, with one "feel[ing] like I do well in physics" [E10]. The other described academic success, "being a sort of star-student in…AP Physics," in high school establishing a strong physics identity foundation, which "has been reinforced by being heavily involved in the physics community at my university" [G8]. The remainder of responses (16 of 18, 88.9%) reflected a negative perception of their abilities to understand physics content and/or perform physics tasks. All participants who wrote about their perceived competence described their abilities negatively. Students reported, "it takes me much longer to understand 'elementary' concepts" [E48] and "college level physics has been pretty difficult for me…Especially [concepts] relating to integrals and derivations without professor help" [B7]. A participant described a lack of knowledge and skills writing, "I don't think I have the necessary systematic approach to everyday problems, nor possess enough intelligence to be considered more of a physics person" [D4]. Many commented on their perceived abilities as "not confident in my classes" [E3], where "the math portion can deter me from wanting to learn more" [E30], leading to "a fear I'm going to start failing all my classes" [D25]. Students did not frame their struggles as part of a growth process, but as fixed realities such as "not fit[ting] the mold of an 'ideal' physics student…struggling academically" [B9].

Students with a low physics identity (A–C) were more than twice as likely to frame their response in terms of performance/competence (5 of 18, 27.8%) as compared with students with mid-to-high physics identities (13 of 102, 12.7%). They were also more likely to describe their performance/competence negatively (100% of A–C responses versus 85% of D-G responses). Students who wrote about performance/competence were not only more likely to have lower physics identity but were also more likely to be upperclassmen: 77.8% (14 of 18), a higher percentage than the overall survey percentage of 59.2% juniors and seniors. A senior noted that "over the years I have become less and less confident in my skills as a physicist" [A2, Senior], and a junior mentioned, "feels like I have to put in more effort than some of my peers to reach the same level of understanding on most concepts" [B7].

#### *2. Personal and/or social identities: Self-image, multiple identities, and belonging*

Just over half (62 of 120, 51.7%) of students explained their physics identities within personal and/or social contexts. They described their self-image, other identities, and perceived role in the physics community.

Students who responded in the context of their personal identities described beliefs they held about themselves and their individual characteristics. Statements that they do or do not see themselves as a physics person were assigned the *self-image* code. Responses illustrated agreement with the measure, as all of those who "don't feel like I identify as a physics person yet" [B4] classified themselves with low physics identity (A–C), and those who saw themselves as "very ingrained with my identity as a woman in physics" [E17] answered with a mid-to-high physics identity (D-G).

Many students recognized physics as one of their *multiple identities*, describing other identities they perceived as outside of and separate from their physics identity. *Multiple identities* was the most frequent code applied to participants' explanations of their physics identities (28.3% of responses) overall. Participants pointed to other interests, disciplines, communities, roles, and aspects of their personalities. Regardless of physics identity (A–G), the explanations were similar. The majority of these responses begin with an expression of interest or connection with physics, yet continue with the qualifying words "but" or "however," and proceed to explain "there is more to me as a person" [F3]. One of the students explained, "I want to be a physicist and I know that becoming a researcher means there is a good portion of you that is identified as such, but at the end of the day, I am still me and I have other hobbies and passions outside of physics" [E9]. Students identified with their other interests as a gymnast, writer, dancer, or artist and with other disciplines including philosophy and film studies. These interests also came with "other communities that define me more as a person" [D19]. Multiple identity responses included other roles such as wife, daughter, and friend. These personal identities were also seen separate from a physics person, described as "many *other* aspects to [their] personality" [D14]. Students with mid-high physics identities, D (11 of 26, 42.3%), E (16 of 49, 32.7%) and F (5 of 17, 29.4%), were most likely to explain their physics identities in terms of multiple identities. We also found that the frequency of this explanation is similar for sophomores (7 of 24, 29.2%), juniors (11 of 37, 29.7%), and seniors (13 of 38, 34.2%), while it represented 3 of 20 (15.0%) of freshman responses.

Students expressed the idea that their physics identities do not mix with other roles or aspects of themselves. Participants wrote explicitly about this separation: "many personal experiences and other parts of my life that do not connect to the physics piece of my identity" [E37] and "there are other identities in my life that are more important to me" [F11]. Physics was cited as "not my entire life" [F5 and F9]. Only one of the 35 students expressed multiple identities as complementary to their physics identity, having "a good balance of extra-curriculars that are outside of physics to keep me balanced" [E12].

The third most common code was *belonging*, applied when explanations described a perception of fitting in or feeling excluded from the physics community. It was a response that spanned years in college or university (freshmen through seniors) and physics identity measures (A–G). A variety of sources led to feelings of inclusion or exclusion. These reasons included societal, social, academic, professional, and psychological factors, and they were not universally positive or negative. For example, academics were equally cited for feeling included in and feeling alienated from the physics community.

When responses coded with belonging were organized by physics identity (A–G) and noted to be positive, negative, or neutral, perceived belonging was related to levels of physics identity. See Fig. 6.

Students whose responses reflected positive belonging described feeling "very connected to physics and the physics department" [E12], having "made a ton of friends through physics" [E35], and found the "support of other women in physics and some amazing professors" [E18]. Professionally, they "know that [they] can make a meaningful contribution to the field" [F3]. Many mention research as a significant influence on feelings of belonging due to experiential learning and finding community outside of the standard coursework and departmental professors: "I have been able to do a lot of cool research, learn about really interesting phenomena, and talk to people…to get a better sense of the physics community outside of my university" [E38]. Another student says that she is a "member of a tightknit research group, and welcomed in the broader departmental research center my lab is a part of." She goes on to cite the value of leadership in that her physics identity was "reinforced by being heavily involved in the physics community at my university as society of physics students leader" [G8].

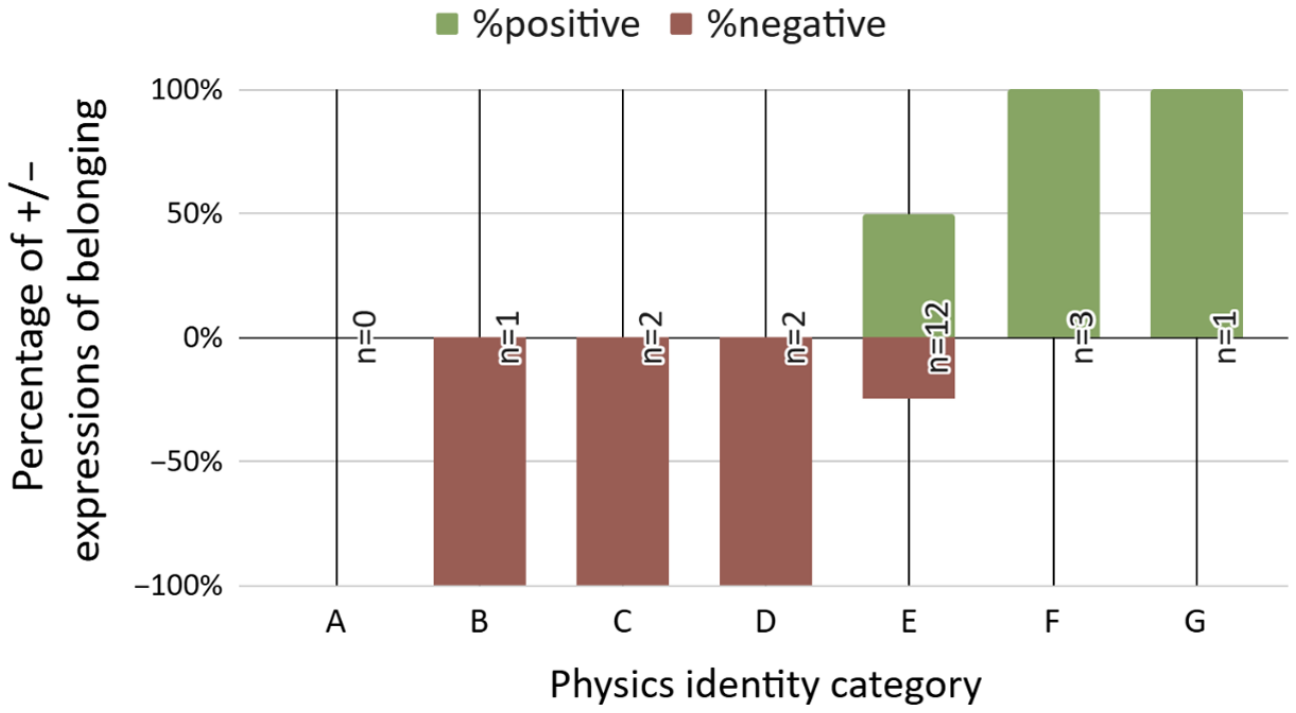

FIG. 6. Students who described their own physics identities in terms of belonging wrote about fitting into (positive expression of belonging) or feeling excluded from (negative expression of belonging) the physics community. Some expressed a mixture of both (neutral expression of belonging). This graph shows the percentage of positive and negative expressions by physics identity category (A is low, G is high) and the number of students in each category who wrote about belonging. The percentage of students who wrote about belonging negatively is represented by the red bars below the axis, and the percentage of students who wrote about belonging positively is represented by the green bars above the axis.

Students felt a lack of belonging for numerous reasons. When parts of their self-identity did not match their perception of a physics person: "I feel like a typical physics person is older than me and usually male…" [D8], it set up a preconceived barrier to acceptance. Academic struggle with parts of the content and/or classes led some students to "feel out of place in my physics courses" [D1]. Others expressed a general lack of connection with peers and the community: "I'm getting an undergrad degree in physics but I feel like I don't know much physics nor do I really feel in touch with [the] physics community" [C6, senior]. Some students wrestled with where they fit professionally in physics, feeling "unsure of what my role in the community would be" [E7], and "not feel[ing] like I belong in physics as a career…even though I love the subject" [E11]. A junior cited exclusionary practices that made her question her abilities and ultimate success: "…there are some physics cliques that can be condescending and make me wonder if I will be able to succeed in physics" [D11].

Self-doubt was present in both positive and negative levels of belonging. A student with negative feelings of belonging did not "feel a part of the community nor do I view myself as someone who should belong there" [B8]. Among students with a higher physics identity who expressed positive senses of belonging, a few still reported low self-efficacy. Although "generally feel[ing] like an accepted and valued part of my community… imposter syndrome also creeps in quite a bit" [E45], and even when one is "pretty involved in my department… I'm not exactly the kind of person you think of when you think about a great physicist" [E36]. These statements indicate that although positive statements of belonging were associated with a higher level of physics identity for students, a positive sense of belonging does not imply self-confidence.

### D. Relationship between student description of a physics person and their own physics identity

To this point, we have presented themes that emerged when we asked participants to describe a physics person, and to explain their own physics identity. In our final cycle of analysis, we compared the responses to these two questions as a whole and for each individual.

Overall, when students explained their own physics identity, they were more likely to use social or personal perspectives than academic ones. They tended to write about having multiple identities and their sense of belonging more than their performance or competence in physics. This was in contrast to their conceptions of a physics person, who was most likely to be described as having excellent academic understanding, engaging in research, and possessing a deep and somewhat singular passion for physics. They described physicists as "be[ing] very centered in only physics and math-based ideas, and to not be very interested in fields outside of physics." [E4]. Only 11 of 120 (9%) participants included a social dimension, "being involved in the community" [D5] or "engag[ing] with others in the field" [D21], when describing a physics person. Of these 11 students, many commented that the involvement and engagement in the physics community were through research.

Interest was the second most common code for both their explanations of their own identity (34 of 120, 28.3%), and their descriptions of a physics person (33 of 120, 27.5%). Common phrases for both sets of responses included enjoyment, love, and passion for physics. Although students were resolute in their own interest in physics when explaining their own identity, they were quick to add negative qualifying statements claiming other identities, academic or research weaknesses, or uncertainty that they fit in. One student wrote, "I enjoy physics but I don't think it's my entire personality, and I am also not amazing at physics" [E27]. In contrast, interest was enough, and was not accompanied with qualifiers or extra requirements, in students' descriptions of a physics person.

The way students depicted academic aptitude in physics was different when they considered themselves versus a physics person. A smaller fraction of students (18 of 120, 15.0%) wrote about their academic abilities than those who included academic understanding in their description of a physics person (36 of 120, 30.0%). Among the students who expressed their physics identity in terms of academic abilities, most described skills they lacked instead of their capabilities, and tended to have lower physics identity. Only two wrote about their skills positively. Twice as many students, distributed across all identity levels (A–G), defined a physics person as having a strong academic understanding of physics. The prevalent notion that a physics person's academic abilities were easy and natural was absent in descriptions of the participants' own identities, where physics was seen as difficult and a struggle.

For each individual student, we examined whether there was a relationship between how they described a physics person and how they described their own physics identity. Nearly half of the students had an explicit relationship, meaning that the language they used to describe a physics person was consistent with their own evaluation. We interpreted this as an indication that they were using the same metric to judge themselves and a physics person. This relationship was positive when a student claimed to have the same qualities with which they attributed to a physics person, and negative when a student cited lacking or being deficient in the qualities they identified with a physics person.

Participants with a positive relationship were more likely to have a higher physics identity. These students expressed their characteristics and habits reflecting the same traits as a physics person. A senior explained her own identity in terms of relative time spent: "A lot of my time is spent doing physics," then described a physics person in a similar way, as "a person who does physics as a majority of their studies or work" [E8]. Another student centered her responses on the physics community and with ideas of belonging. She attributed her physics identity to the fact that she "largely identif[ies] with others in the community" and defined a physics person as "excited to work with people from a variety of backgrounds and perspectives" [F3]. Students with a negative relationship between their responses typically had lower physics identity. Most contrasts were based on the perception that physics people have considerable academic aptitude which comes naturally or easily, which was at odds with their own experiences in learning physics or doing research. One student described her identity through a lack of fitting in and insufficient academic abilities: "I feel I do not fit the mold of the 'ideal' physics student and have been struggling academically." She described a physics person as the exact opposite, as one who "fit[s] within the physics community and [is] successful academically and with research" [B9]. Another participant expressed doubt that they were "passionate enough," when a physics person is "…passionate about your subject" [C1]. A junior described "hav[ing] many other aspects to my personality" separate from physics, but a physics person integrates the subject, "carrying that interest into many aspects of their personal lives" [D14].

A large portion of our participants had no explicit relationship between their responses. In many instances this was due to a lack of elaboration, a limitation of the open-ended survey response method. An example of a pair of responses with no clear relationship is a student who explained her own identity as “When I think of a physicist, I think of myself;” and went on to describe a physics person as “A person who enjoys and respects the field of physics, someone who is always willing to learn about new physics” [G5]. For others, the way they described a physics person was completely unrelated to how they explained their own physics identities. A student who described her own identity in terms of perceived academic abilities: “Sometimes I don’t consider myself to be the smartest to…fit the role of a Physics person,” then went on to describe a physics person in terms of curiosity and diligence, “Having a passion for understanding the universe around you, having a great work ethic” [E16]. Another student with unrelated descriptions defined a physics person in terms of interest, “A physics person enjoys physics and related disciplines (I think that’s the only criterion),” but then described her own physics identity in terms of belonging: “I generally feel like an accepted and valued part of my community, but imposter syndrome also creeps in quite a bit” [E45]. Participants with no clear relationship between their responses were equally likely across all identity levels (A–G).

## V. DISCUSSION

In this study, we examined undergraduate women’s definitions of what it means to be a physics person, as well as descriptions of their own physics identities. Prior studies have analyzed how undergraduate physics students see themselves as physics people without directly asking them to define a physics person [2,16,21,22,41,53]. By explicitly asking this question, it allowed us to understand the image to which undergraduate women are comparing themselves. The relationship between their image of a physics person and how they describe their own physics identity provides insight about where they are in their physics identity development. We found distinct differences in the way students described a physics person and how they explained their own physics identity. Due to the nature of the survey (open-ended narrative responses), we cannot claim to have a full understanding of the students’ definitions of a physics person or a complete description of their own identity. Omission of a certain topic in a student’s response does not imply its absence. Yet we value what these women chose to write as salient to them and their experiences.

If identity is associated with an aspiration to be a certain kind of person [37] or a trajectory toward becoming a core member of a community [39], it is problematic when students describe themselves as lacking the characteristics of the kind of person they aspire to be. Students who conceive the practices, values, and goals of a physics person as different from their own may experience tensions or crises [26,54] in their physics identity development. Our findings suggest three tensions experienced by undergraduate women in physics in the development of their physics identity: (1) the tension between identifying entirely as a physics person and having other identities that are important to who they are, (2) the tension between the need for belonging in the physics community and their view of themselves as a physics person, and (3) the tension between their expectations of a physics person and their beliefs about their own abilities or interest.

### A. The tension between identifying entirely as a physics person and having other identities that are important to who they are

Students explained their own physics identity most frequently by describing their multiple identities. The responses addressed the space outside of and separate from their physics identity (see Figs. 2 and 5), conveying that there are aspects of their identity that are different from a physics person. The tension arises from the conception that a physics person has a singular and deep interest in physics, devoting most of their time and energy to the discipline. When students have other identities and social roles that take time and energy outside of physics, they question whether they can be a physics person given their other identities. In their study of STEM graduate students titled “How can I be me and a scientist?” Tran and colleagues establish that this tension is especially present for many underrepresented populations in STEM, who felt pressured to choose between STEM expectations, such as more time in the lab, and community obligations, such as recruitment or outreach to their ethnic or gender group. Many needed to restructure what it meant to be a STEM person and a member of a minoritized group in order to persist [55]. For women in particular, the tension between identifying as a physics person and having other interests and identities has been shown to have a negative impact on their participation. Ferriman and colleagues found that math and science careers that are perceived as demanding large amounts of dedicated time, with little to no space for multiple interests, have fewer women [56]. Social identity research recognizes that we all have multiple identities, but the social structure surrounding an individual affects which identity is activated or committed to [57]. Our findings, together with past research, suggest that doing work to restructure the image of a physics person from one who is singularly focused on the discipline to one who has multiple aspects to their identity would serve to lessen this tension in women’s physics identity development.

### B. The tension between the need for belonging in the physics community and their view of themselves as a physics person

In our study, many students described their physics identity in terms of their perceived fit within a community. For the 21 students who wrote about belonging, the degree of social connection they felt with their physics peers and department members was related to their level of physics identity. All students who wrote about belonging in a positive way had high physics identities (F–G) (see Fig. 6). They described feelings of connection with friends, and being valued and accepted in their physics communities. Students who wrote about lacking a sense of belonging had lower physics identities (A–D). They wrote about feeling different from their peers and unsure of their place, or out of place, in the physics community. There was no discernible pattern across identity letters of who chose to write about belonging, suggesting that a sense of belonging is important in all levels of physics identity development. We found similar percentages of sophomores (21%), juniors (19%), and seniors (18.4%) chose to describe their physics identity in terms of belonging, compared to 10% of freshmen. In Hazari and colleagues' study of first year and senior women physics students, they found a sense of belonging was statistically significant only for seniors' physics identities [41]. Our data suggests that belonging plays a similarly meaningful role for sophomores and juniors as compared to seniors, and a lesser role for freshmen.

Knowing who has or lacks a sense of belonging in physics, and understanding the factors that encourage or discourage the development of belonging is consequential. Research shows that a sense of belonging predicts academic persistence, particularly for women [28] and African American students [29]. In physics, students with a greater sense of belonging have better academic attitudes and outcomes [30,44]. Studies have also shown that better academic outcomes lead to a greater sense of belonging [27,31]. The relationship between belonging and academic achievement is notable for this study since performance and competence were common considerations for students with lower physics identities, and academic understanding was the most frequent description of a physics person. Our findings highlight the idea that social support structures are just as necessary as academic support for undergraduate physics students' identity development.

When students described a physics person in terms of belonging, it was much less about feeling connected or fitting in, but merely participating and being involved in the physics community through research or teaching. Overall, students wrote less frequently about the social aspects of physics when defining a physics person compared to the descriptions of their own physics identity. The differences in how and how much women chose to describe belonging for themselves and for a physics person may simply reflect the differences between ways of knowing and thinking about one's own needs compared to the needs of others as it is easier to identify one's own sense of belonging than someone else's. However, another possible explanation is derived from Nadelson and colleagues, who propose that progression in STEM professional identity requires less social support scaffolding over time, relying less and less on social cues [34]. Similarly, undergraduate women might not have considered belonging as important for a physics person because they think that STEM professional identity requires less social support as one progresses towards a physics career, garnering more experience and projected expertise. It is unclear if either version of this progression is what our participants were expressing. Regardless, young students toward the beginning of their physics identity development, including all undergraduate years, have a need for belonging that may feel different from the social need experienced by professionals. As Lewis states, "science career attainment is a social process," and "relies on the judgement and invitation of practicing scientists through every phase" [58]. Rosenberg *etal*. noted that a social process that includes engagement in formal and informal leadership and mentorship increases a sense of belonging for undergraduate women [35].

### C. The tension between their expectations of a physics person and their beliefs about their own abilities or interest

Students described a physics person as someone who is good at physics, interested in the subject sometimes to the point of it being a singular dedication, involved in research, or part of the physics community (see Fig. 3). When students reflected on their own identity compared with a physics person, many expressed doubts about their capacity to meet their own expectations of what it means to be a physics person. The uncertainty of whether they could meet the expectations was notable when students wrote about their interest and academic abilities in physics.

Interest was the second most common way women chose to describe a physics person, and to explain their own physics identity. For a physics person, interest was described as enjoyment and enthusiasm for the subject. When students wrote about their own interest, they also expressed it in a positive light. But more than ⅔ of the responses continued with a negative qualifier. They conveyed enjoyment and love for physics, but went on to describe doubts about whether they were smart enough or dedicated enough. This contextualization of interest lends support to Hazari's study of first-year and senior women physics students that found that interest is mediated by performance/competence and recognition [41]. When students are struggling with a subject that they are interested in and enjoy, they question whether their interest is enough to make them a physics person. This perceived space between their own identity, and their image of a physics person may threaten their persistence in physics. Although interest may be necessary to bring women into physics, it might not be enough to keep them there.

The most common way undergraduate women in physics chose to describe a physics person was someone who finds math and solving physics problems to be natural, intuitive, and easy. The frequency of this description indicates the pervasiveness of the natural-genius stereotype throughout undergraduate years and physics identity levels. Nearly all students who wrote about their academic competence and performance in physics when explaining their own physics identity talked about it in a negative way, and were more likely to report lower identity. Again, tension in physics identity development occurs when students' conception of a physics person does not match their own experiences or perceived abilities. This suggests that if physics professionals acknowledge struggle, having to work hard to succeed, or moments of self-doubt in their own journeys in physics, it will benefit the physics identity of undergraduate women in physics.

## VI. LIMITATIONS

The participants in this paper had voluntarily attended a CUWiP. The respondents may have had an above-average affinity for and identification with physics compared to all undergraduate women in physics. However, having a better sense of themselves as a physics person may have enabled participants to provide more articulate and detailed explanations. An additional limitation of the study is that data were collected using an online survey. Follow-on interviews with a subset of survey participants were conducted at a later date to further explore the research questions (see, e.g., Akesson *et al.*, 2025) [48]. The current study also focuses on women in physics and lacks a comparative sample of undergraduate male physics students' responses. It may be informative to compare the relative differences between how men define a physics person and describe their own physics identities as members of the dominant population.

## VII. CONCLUSIONS

This study provides greater contextual understanding of the physics identity development of undergraduate women in physics. Consistent with existing literature, we found that interest, academic performance, and competence play a large role in the development of their physics identity. We also found that perceived social support and identities outside of physics strongly influence physics identity development. Students' conceptions of a physics person provided further insight by describing the model of a physics person to which they were comparing themselves. Our data allowed us to look for trends across the 120 participants while also examining responses for each unique student. Women find challenges in integrating their multiple identities with their physics identity. Students from all undergraduate years have a need to feel part of a social community. When students perceive themselves as "not enough" to meet their expectations of a physics person, it creates tension in their physics identity development.

It is valuable to have women's perceptions of a physics person separate from descriptions of their own physics identity. Women's conceptions of a physics person provide a point of comparison for how they judge themselves and evaluate their trajectory. Our findings inform three implications for undergraduate physics educators, mentors, research group leaders, and undergraduate students themselves.

Working to *normalize multiple identities* within the physics community is important. Students should see that people they deem as experts also have multiple identities and interests outside of physics. When students recognize that being a physics person can coexist with other interests and identities, they may be less apt to question if they are interested or dedicated enough, or feel that physics conflicts with their other identities.

Second, recognizing the importance of *community and social support* in the development of early physics identity may be as important as academic support. Any opportunity to create a sense of community through mentoring or creating spaces for students or students and faculty to come together will serve to benefit undergraduate physics identity. For those in positions to create and organize spaces for social support, it is important to consider that the social needs of students may be greater than those in professional physics positions.

Third, those in senior positions, including faculty, advisors, and research group leaders, should work to *reframe struggle* by validating difficulties and sharing stories of struggle and obstacles that they have overcome. Countering the idea that physics understanding and skills are inherent and that physics comes easily if you are a physics person is important, and can be invaluable for students with lower physics identity.

Our study is limited by its reliance on participant survey responses. Follow-up interviews would help clarify the magnitude of these tensions and reveal potential limitations in the identity measure itself. We used the survey question with circles that illustrate the overlap between their identity and a physics person (Fig. 2) to understand how students characterize their identity. Interviews could help understand how other identities influence the formation of their physics identity, and the degree to which the importance of other identities is an artifact of the survey question design.

Framing our discussion around tensions also opens new directions for future research that differ from prior studies. Existing work on physics identity formation has largely examined the roles of race, gender, family, and culture [17,45,50,55,57]. While this research remains essential, our data suggest the need for a broader perspective across all physics undergraduates. In particular, future work should examine how students who identify as artists, writers, parents, athletes, or members of other groups navigate the integration of those identities with their physics identities. This connects directly to the second and third tensions: belonging and expectations. Identity research has traditionally focused on students and their experiences, but greater attention should also be directed toward the physics discipline itself. Developing a deeper understanding of when and how physics communicates messages about in-groups and out-groups, as well as the expectations associated with being a physicist, may help disrupt the stereotypical view of physics that persists into undergraduate education.

## ACKNOWLEDGMENTS

Thank you to the students who have participated in this work, and our advisory board members. This material is based upon work supported by the National Science Foundation under Grant No. DUE-2044232.

## DATA AVAILABILITY

The data that support the findings of this article are not publicly available because they contain sensitive personal information. The data are available from the authors upon reasonable request.

[1] A. Kelly, A discouraging process: How women are eased out of science, Bull. Sci. Technol. Soc. **3**, 351 (1983).

[2] R. M. Lock and Z. Hazari, Discussing underrepresentation as a means to facilitating female students' physics identity development, Phys. Rev. Phys. Educ. Res. **12**, 020101 (2016).

[3] L. McCullough, Women in physics: A review, Phys. Teach. **40**, 86 (2002).

[4] H. J. Walberg, Physics, femininity, and creativity, Dev. Psychol. **1**, 47 (1969).

[5] M. B. Ruskai, How Stereotypes About Science Affect the Participation of Women (1989).

[6] Newest Data Shows Mixed Progress for Women and Marginalized Groups in Physics Higher Ed, http://www.aps.org/publications/apsnews/202211/newest-data.cfm.

[7] H. Götschel, No space for girliness in physics: Understanding and overcoming the masculinity of physics, Cult. Stud. Sci. Educ. **9**, 531 (2014).

[8] B. A. Adegoke, Impact of interactive engagement on reducing the gender gap in quantum physics learning outcomes among senior secondary school students, Phys. Educ. **47**, 462 (2012).

[9] M. Badri, K. A. Mazroui, A. Al Rashedi, and G. Yang, Variation by gender in Abu Dhabi high school students' interests in physics, J. Sci. Educ. Technol. **25**, 232 (2016).

[10] R. Koul, Work and family identities and engineering identity, J. Eng. Educ. **107**, 219 (2018).

[11] M. Baran, Gender differences in high school students' interests in physics, Asia-Pac. Forum Sci. Learn. Teach. **17**, 6 (2016).

[12] J. Abraham and K. Barker, Exploring gender difference in motivation, engagement and enrolment behaviour of senior secondary physics students in New South Wales, Res. Sci. Educ. **45**, 59 (2015).

[13] S. Cheryan, S. A. Ziegler, A. K. Montoya, and L. Jiang, Why are some STEM fields more gender balanced than others?, Psychol. Bull. **143**, 1 (2017).

[14] M. M. Williams and C. E. George-Jackson, Using and doing science: Gender, self-efficacy, and science identity of undergraduate students in STEM, J. Women Minor. Sci. Eng. **20**, 99 (2014).

[15] H. B. Carlone and A. Johnson, Understanding the science experiences of successful women of color: Science identity as an analytic lens, J. Res. Sci. Teach. **44**, 1187 (2007).

[16] Z. Hazari, G. Sonnert, P. M. Sadler, and M.-C. Shanahan, Connecting high school physics experiences, outcome expectations, physics identity, and physics career choice: A gender study, J. Res. Sci. Teach. **47**, 978 (2010).

[17] L. Avraamidou, Identities in/out of physics and the politics of recognition, J. Res. Sci. Teach. **59**, 58 (2022).

[18] T. Perez, J. G. Cromley, and A. Kaplan, The role of identity development, values, and costs in college STEM retention, J. Educ. Psychol. **106**, 315 (2014).

[19] Elaine Seymour and N. M. Hewitt, *Talking about Leaving: Why Undergraduates Leave the Sciences* (Westview Press, Boulder, Colorado, 1997).

[20] G. Trujillo and K. D. Tanner, Considering the role of affect in learning: Monitoring students' self-efficacy, sense of belonging, and science identity, CBE Life Sci. Educ. **13**, 6 (2014).

[21] Z. Y. Kalender, E. Marshman, C. D. Schunn, T. J. Nokes-Malach, and C. Singh, Why female science, technology, engineering, and mathematics majors do not identify with physics: They do not think others see them that way, Phys. Rev. Phys. Educ. Res. **15**, 020148 (2019).

[22] E. Bottomley, A. Kohnle, K. I. Mavor, P. J. Miles, and V. Wild, The relationship between gender and academic performance in undergraduate physics students: The role of physics identity, perceived recognition, and self-efficacy, Eur. J. Phys. **44**, 025701 (2023).

[23] Z. Y. Kalender, E. Marshman, C. D. Schunn, T. J. Nokes-Malach, and C. Singh, Gendered patterns in the construction of physics identity from motivational factors, Phys. Rev. Phys. Educ. Res. **15**, 020119 (2019).

[24] S. L. Kuchynka, K. Salomon, J. K. Bosson, M. El-Hout, E. Kiebel, C. Cooperman, and R. Toomey, Hostile and

benevolent sexism and college women's STEM outcomes, Psychol. Women Q. **42**, 72 (2018).

[25] R. Dou and H. Cian, Constructing STEM identity: An expanded structural model for STEM identity research, J. Res. Sci. Teach. **59**, 458 (2022).

[26] P. W. Irving and E. C. Sayre, Identity statuses in upper-division physics students, Cult. Stud. Sci. Educ. **11**, 1155 (2016).

[27] G. L. Cohen and J. Garcia, Identity, belonging, and achievement: A model interventions implications, Curr. Dir. Psychol. Sci. **17**, 365 (2008).

[28] C. Good, A. Rattan, and C. S. Dweck, Why do women opt out? Sense of belonging and women's representation in mathematics, J. Pers. Soc. Psychol. **102**, 700 (2012).

[29] L. R. M. Hausmann, J. W. Schofield, and R. L. Woods, Sense of belonging as a predictor of intentions to persist among African American and White first-year college students, Res. High. Educ. **48**, 803 (2007).

[30] S. Cwik and C. Singh, Students' sense of belonging in introductory physics course for bioscience majors predicts their grade, Phys. Rev. Phys. Educ. Res. **18**, 2022

[31] J. D. Edwards, R. S. Barthelemy, and R. F. Frey, Relationship between course-level social belonging (sense of belonging and belonging uncertainty) and academic performance in general chemistry 1, J. Chem. Educ. **99**, 71 (2022).

[32] A. Godwin, The development of a measure of engineering identity, in *ASEE Annual Conference & Exposition* (ASEE, New Orleans, LA, 2016).

[33] T. D. Hudson, K. J. Haley, A. J. Jaeger, A. Mitchall, A. Dinin, and S. B. Dunstan, Becoming a legitimate scientist: Science identity of postdocs in STEM fields, Rev. High. Educ. **41**, 607 (2018).

[34] L. S. Nadelson, S. P. McGuire, K. A. Davis, A. Farid, K. K. Hardy, Y.-C. Hsu, U. Kaiser, R. Nagarajan, and S. Wang, Am I a STEM professional? Documenting STEM student professional identity development, Stud. High. Educ. **42**, 701 (2015).

[35] J. L. Rosenberg, N. Holincheck, K. Fernández, B. W. Dreyfus, F. Wardere, S. Stehle, and T. N. Butler, Role of mentorship, career conceptualization, and leadership in developing women's physics identity and belonging, Phys. Rev. Phys. Educ. Res. **20**, 010114 (2024).

[36] J. E. Stets, P. S. Brenner, P. J. Burke, and R. T. Serpe, The science identity and entering a science occupation, Soc. Sci. Res. **64**, 1 (2017).

[37] J. P. Gee, Chapter 3: Identity as an analytic lens for research in education, Rev. Res. Educ. **25**, 99 (2000).

[38] J. P. Gee, Discourse analysis: What makes it critical?, in *An Introduction to Critical Discourse Analysis in Education*, edited by R. Rogers (Lawrence Earlbaum Associates,, 2004).

[39] E. Wenger, Conceptual tools for CoPs as social learning systems: Boundaries, identity, trajectories and participation, *Social Learning Systems and Communities of Practice*, edited by C. Blackmore (Springer, London, 2010), pp. 125–143.

[40] A. Y. Kim, G. M. Sinatra, and V. Seyranian, Developing a STEM identity among young women: A social identity perspective, Rev. Educ. Res. **88**, 589 (2018).

[41] Z. Hazari, D. Chari, G. Potvin, and E. Brewe, The context dependence of physics identity: Examining the role of performance/competence, recognition, interest, and sense of belonging for lower and upper female physics undergraduates, J. Res. Sci. Teach. **57**, 1583 (2020).

[42] Z. Hazari, P. M. Sadler, and G. Sonnert, The science identity of college students: Exploring the intersection of gender, race, and ethnicity, J. Coll. Sci. Teach. **42**, 82 (2013).

[43] K. L. Lewis, J. G. Stout, S. J. Pollock, N. D. Finkelstein, and T. A. Ito, Fitting in or opting out: A review of key social-psychological factors influencing a sense of belonging for women in physics, Phys. Rev. Phys. Educ. Res. **12**, 020110 (2016).

[44] J. G. Stout, T. A. Ito, N. D. Finkelstein, and S. J. Pollock, How a gender gap in belonging contributes to the gender gap in physics participation, in *AIP Conference Proceedings* (AIP, Philadelphia, PA, USA, 2013), p. 402.

[45] I. R. Johnson, E. S. Pietri, F. Fullilove, and S. Mowrer, Exploring identity-safety cues and allyship among black women students in STEM environments, Psychol. Women Q. **43**, 131 (2019).

[46] L. Archer, J. Moote, B. Francis, J. DeWitt, and L. Yeomans, The "exceptional" physics girl: A sociological analysis of multimethod data from young women aged 10-16 to explore gendered patterns of post-16 participation, Am. Educ. Res. J. **54**, 88 (2017).

[47] G. Potvin *et al.*, Examining the effect of counternarratives about physics on women's physics career intentions, Phys. Rev. Phys. Educ. Res. **19**, 010126 (2023).

[48] L. M. Akesson, J. L. Rosenberg, B. W. Dreyfus, and N. M. Holincheck, Physics identity: Examining the role of recognition for undergraduate women in physics, in *2025 PERC Proceedings* (AAPT, Washington, DC, 2025), pp. 22–27).

[49] J. N. Tripp and X. Liu, Towards defining STEM professional identity: A qualitative survey study, J. STEM Educ. Res. **8**, 348 (2025).

[50] M. Ong, J. M. Smith, and L. T. Ko, Counterspaces for women of color in STEM higher education: Marginal and central spaces for persistence and success, J. Res. Sci. Teach. **55**, 206 (2018).

[51] M. M. McDonald, V. Zeigler-Hill, J. K. Vrabel, and M. Escobar, A single-item measure for assessing STEM identity, Front. Educ. **4**, 78 (2019).

[52] J. Saldaña, The Coding Manual for Qualitative Researchers (2021).

[53] S. Hyater-Adams, C. Fracchiolla, T. Williams, N. Finkelstein, and K. Hinko, Deconstructing Black physics identity: Linking individual and social constructs using the critical physics identity framework, Phys. Rev. Phys. Educ. Res. **15**, 020115 (2019).

[54] J. E. Marcia, Development and validation of ego-identity status, J. Pers. Soc. Psychol. **3**, 551 (1966).

[55] M. Tran, F. A. Herrera, and J. Gasiewski, STEM graduate students' multiple identities: How can I be Me and a

scientist, in (National Association of Research on Science Teaching, Orlando, FL, 2011).

[56] K. Ferriman, D. Lubinski, and C. P. Benbow, Work preferences, life values, and personal views of top math/science graduate students and the profoundly gifted: Developmental changes and gender differences during emerging adulthood and parenthood, J. Pers. Soc. Psychol. **97**, 517 (2009).

[57] P. J. Burke, Relationships among multiple identities, in *Advances in Identity Theory and Research*, edited by P. J. Burke, T. J. Owens, R. T. Serpe, and P. A. Thoits (Springer US, Boston, 2003), pp. 195–214.

[58] B. F. Lewis, A critique of literature on the underrepresentation of African Americans in science: Directions for future research, J. Women Minorities Sci. Eng. **9**, 14 (2003).